\documentclass[pre,twocolumn,aps,superscriptaddress]{revtex4-2}
\usepackage[T1]{fontenc}
\usepackage[latin9]{inputenc}
\usepackage{array}
\usepackage[pdftex]{graphicx}
\usepackage{mathrsfs}
\usepackage{multirow}
\usepackage{amsmath}
\usepackage{amsthm}
\usepackage{amssymb}
\usepackage{mathbbol}
\usepackage{graphicx}
\usepackage{latexsym}
\usepackage{textcomp}
\usepackage{mathtools}
\usepackage{amsmath}
\usepackage{xcolor}
\usepackage{braket}
\usepackage{physics}

\usepackage[pdftex,colorlinks=true,urlcolor=blue,linkcolor=blue,citecolor=blue,breaklinks=true,bookmarks=false,hypertexnames=false]{hyperref}
\DeclareMathAlphabet\mathbfcal{OMS}{cmsy}{b}{n}
\usepackage{bm}
\definecolor{mycolor1}{rgb}{0.1, 0.6, 0.6}
\usepackage{calligra} \DeclareMathAlphabet{\mathcalligra}{T1}{calligra}{m}{n} \DeclareFontShape{T1}{calligra}{m}{n}{<->s*[2.2]callig15}{}

\newcommand{\push}{\hspace{0.05cm}}
\newcommand{\pull}{\hspace{-0.05cm}}

\begin{document}

\title{Density-and-phase domain walls in a condensate with dynamical gauge potentials}

\author{Sayak Bhattacharjee}
\email{sayakb@iitk.ac.in}
\affiliation{Department of Physics, Indian Institute of Technology Kanpur, Kanpur 208016, India}
\affiliation{Max Planck Institute for the Physics of Complex Systems, Dresden 01187, Germany}
\author{Roderich Moessner}
\affiliation{Max Planck Institute for the Physics of Complex Systems, Dresden 01187, Germany}
\author{Shovan Dutta}
\affiliation{Max Planck Institute for the Physics of Complex Systems, Dresden 01187, Germany}
\date{\today}

\begin{abstract}
We show how one can generate domain walls that separate high- and low-density regions with opposite momenta in the ground state of a harmonically trapped Bose-Einstein condensate using a density-dependent gauge potential. Within a Gross-Pitaevskii framework, we elucidate the distinct roles of vector and scalar potentials and how they lead to synthetic electromagnetic fields that are localized at the domain wall. In particular, the kinetic energy cost of a steep density gradient is compensated by an electrostatic field that pushes particles away from a special value of density. We show numerically in one dimension that such a domain wall is more prominent for repulsive contact interactions, and becomes metastable at strong electric fields through a first-order phase transition that ends at a critical point as the field is reduced. Our findings build on recent experimental developments and may be realized with cold atoms in a shaken optical lattice, providing insights into collective phenomena arising from dynamical gauge fields.
\end{abstract}

\maketitle

\section{\label{sec:intro}Introduction}

An important challenge facing engineered quantum systems is simulating the physics of gauge theories \cite{zohar2015quantum, wiese2013ultracold, dalmonte2016lattice}. The goal is to probe phenomena such as confinement \cite{tan2021domain} as well as uncover new collective effects. A major advance was to realize artificial magnetic fields for neutral atoms and photons \cite{aidelsburger2018artificial}, which enabled, e.g., a realization of the Haldane model \cite{jotzu2014experimental} and photonic Laughlin states \cite{clark2020observation}. Recent years have seen a coordinated effort to make such fields \emph{dynamical} in order to probe interacting matter and gauge fields \cite{gorg2019realization, schweizer2019floquet}, as in quantum chromodynamics \cite{wiese2014towards, banuls2020simulating}. In particular, density-dependent gauge potentials, which play a key role in Chern-Simons physics \cite{valenti2020synthetic}, have been realized in Bose-Einstein condensates (BECs) by shaking \cite{clark2018observation, yao2022domain} and Raman dressing \cite{edmonds2013simulating, frolian2022realizing}. These gauge fields do not have an independent degree of freedom but already produce intriguing domain walls in the experimental ground state \cite{yao2022domain}, whose generation and dynamics are not well understood. Here, we elucidate how the emergent Lorentz forces allow one to stabilize and engineer a wider class of such domain walls.

Domain walls generally arise as topological defects in nonlinear media, that are of fundamental interest in magnetism \cite{nataf2020domain, evans2020domains} and astroparticle physics \cite{vilenkin1994cosmic, saikawa2017review}, and have applications in information processing \cite{catalan2012domain} and optical communication \cite{song2019recent}. In quantum-gas experiments, domain walls spontaneously excited in quenches provided an important test of Kibble-Zurek universality \cite{clark2016universal, keesling2019quantum}. They are also created deterministically by shining light on parts of a condensate to imprint a phase, leading to dark solitons \cite{burger1999dark}, or by applying a nonuniform magnetic field to a spinor condensate, which led to the observation of Dirac monopoles \cite{ray2014observation} and knot solitons \cite{hall2016tying} (see Ref.~\cite{kengne2021spatiotemporal} for a review). A density-dependent gauge potential, on the other hand, allows one to shape the \emph{ground-state} phase profile by coupling it to the local density \cite{buggy2020hydrodynamics}. This scheme was used by Yao {\it et al} \cite{yao2022domain} to create phase domains in a harmonic trap, where the condensate switches between equal and opposite (canonical) momenta of a double well. However, the density profile itself was unaffected by the potential, i.e., no feedback was observed, limiting the range of accessible physics.

We show that the experimental scenario effectively corresponds to the case of a pure vector potential, for which the Lorentz forces vanish in a static condensate. Generically, applying a density-dependent tilt $\mathbf{A}(\rho).\mathbf{p}$ to the single-particle dispersion $\varepsilon(\mathbf{p})$, as in Ref.~\cite{yao2022domain}, yields both \textit{electric} and vector potentials, which can be used to tailor density as well as phase variations. In particular, we show that instead of switching between opposite momenta, if one lets the single-particle ground state interpolate \emph{smoothly} with density, e.g., by tilting a quadratic dispersion, then the electric potential can give rise to domain walls where the density falls sharply as the phase gradient reverses direction. We discuss a minimal model where such domain walls are tunable over a wide range of experimental parameters. Crucially, they represent ground-state topological structures where the synthetic electromagnetic fields are concentrated and may host previously unknown collective modes. Furthermore, at sufficiently strong fields it can become energetically favorable to annihilate the domain wall through a first-order transition that ends at a critical point, which may be used to probe defect generation by false-vacuum decay relevant to inflationary cosmology \cite{coleman1977fate}.

Below we first discuss the equations of motion within a Gross-Pitaevskii formalism before presenting numerical results for a one-dimensional (1D) model and discussing possible experimental realizations.

\section{\label{sec:eom}Synthetic Lorentz forces}

The hydrodynamic equations of a BEC subject to a dynamical gauge potential were derived in Ref.~\cite{buggy2020hydrodynamics}. Here we explain how the resulting density-dependent electromagnetic forces shape the ground state.

We consider identical bosons with quadratic dispersion and unit mass, i.e., $\varepsilon(\mathbf{p}) = |\mathbf{p}|^2/2$. A synthetic vector potential $\mathbfcal{A}$ shifts the canonical momentum $\mathbf{p} \equiv -{\rm i} \hbar \bm{\nabla}$, rotating the phase of the wave function like a true vector potential acting on a unit charge. This shift results in the kinetic energy $\varepsilon_{\text{kin}} = |\mathbf{p} - \mathbfcal{A}|^2/2$, where $\mathbf{p} - \mathbfcal{A}$ is the mechanical momentum. In shaking experiments a gauge potential may be realized by tilting the dispersion by $\mathbfcal{A} \cdot \mathbf{p}$ \cite{yao2022domain}. However, this does not account for the $|\mathbfcal{A}|^2$ term in $\varepsilon_{\text{kin}}$. Stated differently, such a tilt is equivalent to a vector potential $\mathbfcal{A}$ \emph{and} a scalar potential $- |\mathbfcal{A}|^2/2$. As we show below, these two play very separate roles in a static condensate, with important consequences.

To keep the discussion general, we consider a BEC with arbitrary density-dependent vector and scalar potentials $\mathbfcal{A}(\rho)$ and $\mathcal{B}(\rho)$, respectively. Additionally, the particles are trapped in an external potential $V(\mathbf{r})$ and have pairwise $s$-wave contact interactions of strength $g$, which are both tunable in cold-atom setups \cite{schafer2020tools}. At the mean-field level \cite{pethick2008bose}, the condensate is governed by the Hamiltonian
\begin{equation}
    H = (1/2) |\mathbf{p} - \mathbfcal{A}(\rho)|^2 + \mathcal{B}(\rho) + V(\mathbf{r}) + (g/2) \rho \;.
    \label{eq:hamil}
\end{equation}
The total energy is $\mathcal{E} = \langle \psi | H | \psi \rangle$, where $\psi(\mathbf{r},t)$ is the condensate wave function varying in position $\mathbf{r}$ and time~$t$. Writing $\smash{\psi = \sqrt{\rho} \push e^{{\rm i} \varphi}}$, where $\varphi$ is the phase, we find
\begin{equation}
    \mathcal{E} = 
    \int \! {\rm d}\mathbf{r} \left[ 
    \frac{\hbar^2}{2} \frac{|\bm{\nabla} \rho|^2}{4\rho} 
    + \frac{1}{2} \rho |\mathbf{v}|^2 
    + (\mathcal{B} + V) \rho
    + \frac{g}{2} \rho^2
    \right] \push ,
    \label{eq:energy}
\end{equation}
where $\mathbf{v} \coloneqq \hbar \bm{\nabla} \varphi - \mathbfcal{A}$ is the velocity of the condensate, and $\mathcal{B}(\rho) + V(\mathbf{r}) \coloneqq \mathcal{V}(\rho,\mathbf{r})$ is the net scalar potential. In Eq.~\eqref{eq:energy} the second term gives the classical kinetic energy and the first term describes a quantum correction, which vanishes for $\hbar \to 0$. The third and fourth terms represent potential and interaction energies, respectively.

The equation of motion can be obtained by minimizing the action $S = \int \pull {\rm d}t \push \langle \psi | {\rm i} \hbar \partial_t - H | \psi \rangle$ \cite{kramer1980geometry} with respect to $\rho$ and $\varphi$, with the constraint $\int \pull {\rm d}\mathbf{r} \push \rho(\mathbf{r}) = N$, where $N$ is the total particle number. Using $\langle \psi | {\rm i} \partial_t | \psi \rangle = - \int \pull {\rm d}\mathbf{r} \push \rho \partial_t \varphi$ and Eq.~\eqref{eq:energy} gives the Euler-Lagrange equations
\begin{subequations}
\begin{align}
    \partial_t \rho & + \bm{\nabla} \cdot \mathbf{j} = 0 \;, 
    \label{eq:continuity} \\
    \hbar \partial_t \varphi & 
    + Q + |\mathbf{v}|^2/2 + \Phi + V + g \rho - \mu = 0 \;, 
    \label{eq:hamiltonjacobi}
\end{align}
\end{subequations}
where we have introduced a chemical potential $\mu$ as a Lagrange multiplier for the particle-number constraint, $\mathbf{j} \coloneqq \rho \mathbf{v}$ is the current density, $Q \coloneqq -(\hbar^2/2)(\nabla^2 \sqrt{\rho} / \sqrt{\rho})$ is a quantum potential, and
\begin{equation}
    \Phi \coloneqq \partial_{\rho} (\rho \mathcal{B}) - \mathbf{j} \cdot \partial_{\rho} \mathbfcal{A}     \label{eq:eletricpotential}
\end{equation}
is a potential resulting from the density-dependent fields. Equation \eqref{eq:continuity} is the continuity equation and Eq.~\eqref{eq:hamiltonjacobi} is a quantum Hamilton-Jacobi equation \cite{ballentine2014quantum}, which differs from those of a standard condensate only by the presence of $\Phi$. Note when $\mathbfcal{A}$ and $\mathcal{B}$ do not depend on $\rho$, $\Phi + V$ simply gives the net external potential $\mathcal{V}(\mathbf{r})$. To interpret $\Phi$ generally, we take the gradient of Eq.~\eqref{eq:hamiltonjacobi} to find the Cauchy momentum equation
\begin{equation}
    \frac{D \mathbf{v}}{Dt} = 
    - \bm{\nabla} \left( Q + V + g \rho \right) 
    + \mathbf{E} + \mathbf{v} \cross \mathbf{B} \;,
    \label{eq:cauchy}
\end{equation}
where $D/Dt \coloneqq \partial_t + \mathbf{v} \cdot \bm{\nabla}$ is the convective or total time derivative for a fluid element, and
\begin{equation}
    \mathbf{E} = -\bm{\nabla} \Phi - \partial_t \mathbfcal{A}
    \quad \text{and} \quad 
    \mathbf{B} = -\bm{\nabla} \cross \mathbf{v}
    \label{eq:EMfields}
\end{equation}
are the synthetic electric and magnetic fields, which encapsulate the effects of the density-dependent potentials. Thus, $\Phi$ acts as the electric potential. From Eq.~\eqref{eq:EMfields} the magnetic field is set by the local vorticity and can be rewritten as $\mathbf{B} = \bm{\nabla} \cross \mathbfcal{A} - \hbar \bm{\nabla} \cross \bm{\nabla} \varphi$. The second term vanishes except where $\bm{\nabla} \varphi$ is singular, e.g., at centers of quantised vortices \cite{fetter2001vortices}. Conversely, from Eqs.~\eqref{eq:eletricpotential} and \eqref{eq:EMfields} the electric field is set by both $\rho$ and $\mathbf{v}$.

Note that for $\mathcal{B} = 0$ the Lorentz forces vanish whenever the condensate is stationary. Thus, a nonzero scalar potential is necessary in order to modify the stationary density profiles, including that of any 1D ground state.

For such stationary states, $\mathbf{v}=0$ implies $\bm{\nabla} \varphi = \mathbfcal{A}/\hbar$, i.e., the phase gradient is determined by the local vector potential, which was utilized in Ref.~\cite{yao2022domain} to create phase domains. On the other hand, Eq.~\eqref{eq:hamiltonjacobi} gives a generalized Gross-Pitaevskii equation for the density, 
\begin{equation}
    Q[\rho] + \Phi_0[\rho] + V(\mathbf{r}) + g \rho = 
    \mu - \hbar \omega \;,
    \label{eq:stationarydensity}
\end{equation}
where $\omega$ is the rate of phase winding, which can be different for ground and excited states, and $\Phi_0 \coloneqq \partial_{\rho} (\rho \mathcal{B})$ is the electrostatic potential, which does not depend on $\mathbfcal{A}$. Hence, the roles of the vector and scalar potentials are uncoupled: $\mathcal{B}(\rho)$ changes the density variation caused by the trap, and $\mathbfcal{A}(\rho)$ sets the phase profile.

To understand how the form of $\mathcal{B}$ affects the ground state in particular, 
note that in Eq.~\eqref{eq:energy} it adds an energy per unit volume of $\rho \mathcal{B}(\rho)$, favoring more weight in values of density for which $\rho \mathcal{B}(\rho)$ is reduced. In particular, if the energy cost rises sharply around a special density $\rho_c$, the particles will be pushed away from $\rho_c$ in both directions along the density axis by the electric field, which can give rise to domain walls separating high- and low-density regions, as we illustrate in the next section.

\section{\label{sec:model}Model with domain wall}

We focus on the case $\mathcal{B} = -|\mathbfcal{A}|^2/2$ which is realized by applying only a tilt $\mathbfcal{A}({\rho}) \cdot \mathbf{p}$, as we explained in Sec.~\ref{sec:eom}. For this condition, the density and phase domains will coincide. However, this is not essential and more general profiles may be created by tuning $\mathbfcal{A}$ and $\mathcal{B}$ separately.

\begin{figure*}
    \includegraphics[width=\textwidth]{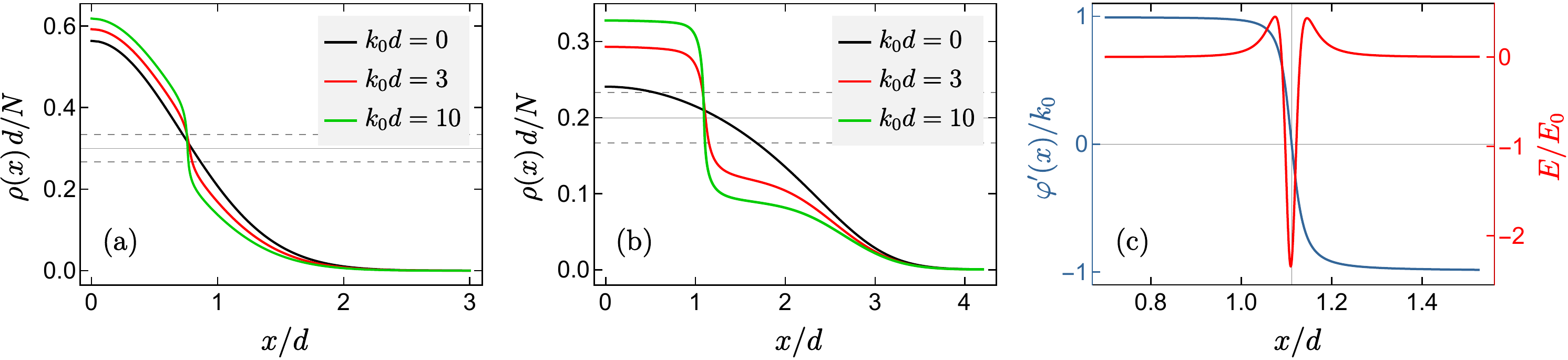}
    \centering
    \caption{(a,b) Ground-state density profiles for $x>0$ of a 1D BEC with $N$ bosons in a harmonic trap of length $d$ with interaction strength $g$ in the presence of a density-dependent gauge potential given by Eq.~\eqref{eq:tanh} with $N l / d = 30$ for (a) $\tilde{g} = 0$ and (b) $\tilde{g} = 40$, where $\tilde{g} \coloneqq (2 N d / \hbar^2) g$ and $k_0 \coloneqq p_0/\hbar$. The solid and dashed horizontal lines correspond to $\rho_c$ and $\rho_c \pm 1/l$, respectively. As the gauge potential is increased a steep slope emerges where $\rho = \rho_c$, becoming more prominent for stronger repulsive interactions. (c) Reversal of the phase gradient (dark blue) and a synthetic, localized electrostatic field [Eq.~\eqref{eq:EMfields}] (red) for the $k_0 d = 3$ curve in (b), where $E_0 \coloneqq \hbar^2 k_0^3 N l / d$. The vertical and horizontal lines show where $\rho = \rho_c$ and $\varphi^{\prime} = 0$, respectively.
    }
    \label{fig:profiles}
\end{figure*}

\subsection{\label{sec:physicalcons}Physical considerations}

The simplest way to create a phase domain wall is by having $\mathbfcal{A}(\rho)$ switch direction depending on whether the local density is above or below $\rho_c$, $\mathbfcal{A} = p_0 \; \text{sign}(\rho - \rho_c) \hat{\mathbf{x}}$, where $p_0$ is the amplitude, $\hat{\mathbf{x}}$ is a unit vector, and $\text{sign}(.)$ is the sign function. In the ground state $\bm{\nabla} \varphi$ follows $\mathbfcal{A}$ to minimize the kinetic energy, i.e., the canonical momentum also changes sign where $\rho$ crosses $\rho_c$ \cite{yao2022domain}. However, this choice gives $\mathcal{B} = -p_0^2/2$, which is simply a constant and does not affect the density profile.

To produce a sharp fall in density the crossover between $\pm p_0$ needs to occur over a finite density interval $l^{-1}$, as exemplified by
\begin{equation}
    \mathbfcal{A} = p_0 \tanh[(\rho - \rho_c) l] \push \hat{\mathbf{x}} \;.
    \label{eq:tanh}
\end{equation}
Note that $l$ has the dimension of volume, reducing to a length in 1D. Then $\mathcal{B} = -(p_0^2/2) \tanh^2[(\rho - \rho_c) l]$ is peaked at $\rho_c$ and penalizes densities in the range $\rho_{\pm} \coloneqq \rho_c \pm 1/l$. For sufficiently large $p_0$ this effect can overcome the kinetic energy cost of steep density gradients [Eq.~\eqref{eq:energy}] and stabilize a domain wall where $\rho$ falls from $\rho_+$ to $\rho_-$. Concurrently, $\bm{\nabla} \varphi$ also changes direction across the domain wall [Eq.~\eqref{eq:tanh}], so the density and phase variations are correlated. For $l \to \infty$, $\mathbfcal{A}$ goes back to the sign function, whereas for $l=0$ the potentials vanish. Thus, a nonzero and finite value of $l$ is necessary to see this physics.

Since $\mathbfcal{A}$ and $\mathcal{B}$ vary appreciably only across the domain wall, the electromagnetic fields in Eq.~\eqref{eq:EMfields} would also be concentrated there. This structure is reminiscent of flux-attached particles that give anyons in fractional quantum Hall physics \cite{valenti2020synthetic, ezawa2008quantum}. Here also it is plausible that the domain walls will have interesting particle-like degrees of freedom, as suggested by first experiments \cite{yao2022domain}.

The $\tanh$ form in Eq.~\eqref{eq:tanh} is by no means a prerequisite. In fact, in Appendix \ref{appsec:piecewiselin} we construct a family of smooth curves that approach a piecewise linear form of $\mathbfcal{A}(\rho)$ and produce even sharper domain walls in 1D.

The density gradient at such a domain wall can be estimated for strong fields from a competition between the electrostatic and kinetic energies. For this purpose, we assume a domain wall of width $w$ across which the density changes by $\Delta \rho \sim 2/l$. The domain wall has a surface area $A$ and a volume $w A$. From Eq.~\eqref{eq:energy} the electrostatic energy cost of having particles in this volume is $\mathcal{E}_{\text{el}} \approx (p_0^2/2) \rho_c w A$. On the other hand, the kinetic energy cost of having a steep gradient of magnitude $s \approx \Delta \rho / w$ is $\mathcal{E}_{\text{kin}} \approx (\hbar^2/2) [s^2 / (4 \rho_c)] w A$. Hence, the net energy cost, with $p_0 \coloneqq \hbar k_0$, is
\begin{equation}
    \mathcal{E}_{\text{dw}} \approx \frac{\hbar^2}{2} A \Delta \rho 
    \left( \frac{s}{4 \rho_c} + \frac{k_0^2 \rho_c}{s} \right) \push ,
    \label{eq:domainwallenergy}
\end{equation}
which is minimized for $s = 2 k_0 \rho_c$. Including the interaction $g$ gives a correction $\sim O(1/k_0)$ to $s$. This estimate agrees very well with numerical simulations in 1D (see Appendix \ref{appsec:slope}). Thus, whereas the density drop is set by $l$, the slope is set by $k_0 \rho_c$ for sufficiently large $k_0$.

\subsection{\label{sec:profiles}Numerical profiles}

To reduce computational cost we explore ground states in 1D, where $\mathbf{v}=0$, which already exhibit the salient features. Such 1D condensates have been realized in highly elongated traps \cite{gorlitz2001realization} where the transverse motion is frozen out, and the interaction $g$ is renormalized \cite{olshanii1998atomic}. We assume that the vector potential in Eq.~\eqref{eq:tanh} points along the longitudinal direction which has a harmonic confinement of frequency $\omega$. We take the trap length $d \coloneqq \sqrt{\hbar/\omega}$ as our unit of length, and rescale the density by the particle number $N$, which gives the dimensionless parameters $\tilde{k}_0 \coloneqq k_0 d$, $\tilde{\rho}_c \coloneqq \rho_c d / N$, $\tilde{l} \coloneqq N l / d$, and $\tilde{g} \coloneqq (2Nd/\hbar^2) g$. From Eq.~\eqref{eq:energy} the rescaled energy functional is given by
\begin{equation}
    \tilde{\mathcal{E}} = 
    \int_{-\infty}^{\infty} \! {\rm d}\tilde{x}  
    \left[ 
    \frac{(\partial_{\tilde{x}} \tilde{\rho})^2}{4\tilde{\rho}} 
    + \tilde{\rho} \mathcal{\tilde{B}}(\tilde{\rho}) 
    + \tilde{x}^2 \tilde{\rho} 
    + \frac{\tilde{g}}{2} \tilde{\rho}^2
    \right] 
    \push ,
    \label{eq:rescaledenergy}
\end{equation}
where $\tilde{\mathcal{E}} \coloneqq 2\mathcal{E}/(N \hbar \omega)$, $\tilde{x} \coloneqq x/d$, $\tilde{\mathcal{B}} \coloneqq 2\mathcal{B} / (\hbar\omega)$, and the rescaled density $\tilde{\rho} \coloneqq \rho d / N$ satisfies $\int \pull {\rm d}\tilde{x} \push \tilde{\rho}(\tilde{x}) = 1$. We minimize $\smash{\tilde{\mathcal{E}}}$ subject to this constraint, using an adaptive grid to accurately resolve the domain walls.

Figure~\ref{fig:profiles}(a) shows the density profiles for $g=0$. When the scalar potential $\mathcal{B}$ is absent this is simply the Gaussian ground state of a harmonic trap. As $k_0$ is increased, a steep slope develops where $\rho$ crosses $\rho_c$, signifying the domain wall. This becomes much more prominent if one turns on repulsive contact interactions, $g>0$ [Fig.~\ref{fig:profiles}(b)]. As seen from Eq.~\eqref{eq:rescaledenergy} such interactions penalize density fluctuations, favoring a uniform profile. For $k_0 = 0$ this effect competes with the trap and leads to a parabolic Thomas-Fermi profile for $\tilde{g} \gg 1$. On the other hand, when a domain wall is established by large $k_0$ the effect of $g$ is to flatten the density on both sides of the wall, producing a wedding cake-like structure.

Figure~\ref{fig:profiles}(c) shows how the phase reverses slope and the synthetic electric field is strongly localized at such a domain wall, as we argued previously. From Eqs.~\eqref{eq:eletricpotential} and \eqref{eq:EMfields} the maximum value of the electric field scales as $k_0^3 l$. Note there is no magnetic field in 1D.

\subsection{\label{sec:phasetran}Discontinuous phase transition}

\begin{figure}
    \includegraphics[width=0.9\columnwidth]{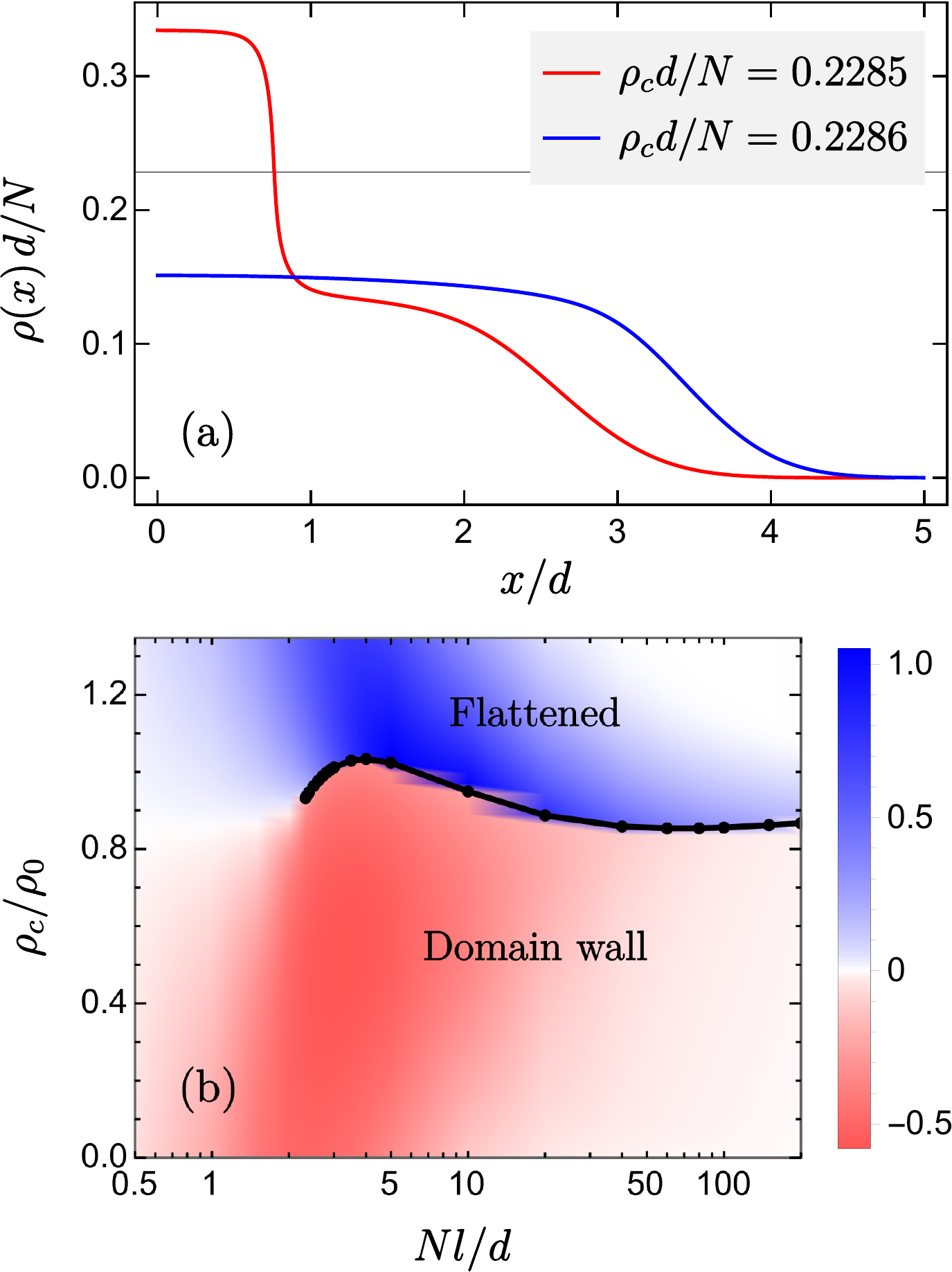}
    \centering
    \caption{(a) Discontinuous transition in the ground-state density for $\tilde{g} = 40$, $N l / d = 30$, and $k_0 d = 5$. As $\rho_c$ crosses above the transition point (horizontal line), it becomes energetically favorable to annihilate the domain structure (red) and create a flatter profile (blue) below $\rho_c$. (b) Phase diagram for $\tilde{g} = 10$ and $k_0 d = 2$, where $\rho_0$ is the peak density for $k_0=0$. The phase boundary (black curve) ends at a critical point for small $l$ where the electric field is weak. The color tracks the potential energy difference, $\mathcal{E}_{\text{pot}} / \mathcal{E}_{0,\text{pot}} - 1$, where $\mathcal{E}_{0,\text{pot}}$ is the potential energy for $k_0 = 0$, showing one approaches the unperturbed ground state for $l \to 0$ and $l \to \infty$. 
    }
    \label{fig:phasetran}
\end{figure}

Creating a domain wall is one way to save electrostatic energy by removing particles from the range $\rho \sim \rho_c \pm 1/l$. Another way is to push the density everywhere below $\rho_c - 1/l$ [see Fig.~\ref{fig:phasetran}(a)]. Such a state also lowers kinetic energy as it is flatter. However, it has high potential energy, as the cloud extends much farther from the center of the trap. This is particularly costly for $\rho_c \ll \rho_0$, where $\rho_0$ is the peak density without the gauge potential. Thus, for small $\rho_c$ a domain wall is energetically favorable. However, as $\rho_c$ is increased beyond $\rho_0$ the flatter state has to become the ground state. For sufficiently strong electric fields the two states are always separated by an energy barrier, at least in the Gross-Pitaevskii formalism, which results in a discontinuous phase transition as shown in Fig.~\ref{fig:phasetran}(b). As one crosses the transition curve the ground state changes dramatically [Fig.~\ref{fig:phasetran}(a)] and the domain wall becomes metastable.

The decay of such a metastable state or ``false vacuum'' through quantum fluctuations plays a key role in models of the early universe \cite{coleman1977fate}, and experimental efforts are underway to probe this physics with quantum simulators \cite{abel2021quantum, song2022realizing}. The metastable lifetime depends on the energy barrier, which in our model can be tuned continuously by the gauge potential. In fact, as the electric field is reduced by decreasing $l$, we find the energy barrier shrinks to zero as the transition curve ends at a critical point [Fig.~\ref{fig:phasetran}(b)]. For smaller values of $l$ the two states are described by the same energy minimum and are no longer distinguishable. This structure is similar to the liquid-gas phase transition of water. At the critical point, the ground-state observables (e.g., the central density) vary infinitely fast with the system parameters (see Appendix \ref{appsec:criticalpoint}), as in a continuous phase transition.

Note that for $l \to 0$ or $l \to \infty$ the scalar potential $\mathcal{B}$ becomes insignificant and the ground state approaches that of the unperturbed system, as seen in Fig.~\ref{fig:phasetran}(b).

\subsection{\label{sec:realization}Experimental realization}

The key physical ingredient in our setup is that the minimum of the single-particle dispersion varies from $-k_0$ to $+k_0$ as the local density changes over a finite interval where the domain wall would appear. For this purpose we assumed a quadratic dispersion and a tilt that is a nonlinear function of density, saturating at $\pm k_0$ [e.g., as in Eq.~\eqref{eq:tanh}]. Such a nonlinear dependence may be hard to realize in experiments. However, one can circumvent the problem by turning on a lattice in the $x$ direction, where the quasimomentum has a natural cutoff given by the Brillouin zone boundary, which could act as $k_0$. Then one requires only a linear tilt $\mathbfcal{A} = A_0 (\rho - \rho_c) \hat{\mathbf{x}}$, where $A_0$ controls the strength of the synthetic electrostatic field. This linear tilt was already implemented in Ref.~\cite{yao2022domain} by shaking an optical lattice and oscillating the interaction strength $g(t)$ synchronously with the micromotion; depending on whether the occupation of a quasimomentum is in or out of phase with $g$, it gains or loses an average energy in the stroboscopic Hamiltonian. As $g$ can be varied over a wide range through a Feshbach resonance \cite{chin2010feshbach}, it is plausible that one can realize a sharp domain wall in density as well as probe its metastability and hysteresis across the discontinuous phase transition.

\section{\label{sec:conclusions}Summary and outlook}

We have shown that a matter-dependent gauge potential can give rise to sharp domain walls in a BEC where synthetic electromagnetic fields are localized. A domain wall in the ground-state density is stabilized by the electric force which in turn originates from the density variation in a trap. Such a domain wall may be realized with cold atoms in a shaken lattice where one can probe its stability across a tunable first-order phase transition. 

Our findings motivate several open questions for future studies. First, how does one understand the dynamics of the domain walls? Already for the usual Gross-Pitaevskii equation, solitonic excitations exhibit rich dynamics \cite{kengne2021spatiotemporal}. What new degrees of freedom are introduced by the localized electromagnetic fields? How do velocity-dependent electric forces [Eq.~\eqref{eq:eletricpotential}] and the associated lack of immediate Galilean invariance \cite{buggy2020gauge} manifest themselves in these dynamics? Does the domain wall behave like a particle with a negative charge-to-mass ratio, as suggested experimentally \cite{yao2022domain}? Another question of fundamental interest is what additional structures emerge in higher dimensions. This is particularly appealing because starting in two dimensions a BEC can have vortices in the ground state \cite{lin2009synthetic} and a density-dependent magnetic force in Eq.~\eqref{eq:cauchy}, which add more nonlinearity to the problem and may alter the stability of the domain walls \cite{malomed2019vortex}. Recent advances in experimental capabilities provide a great incentive to answer these questions, which will be crucial to develop our understanding of collective structures arising from coupled matter and gauge fields.

\acknowledgements

SB gratefully acknowledges the Max Planck Summer Internship Fellowship 2022 for support during his stay at MPIPKS, where the work was carried out. This work was in part supported by the Deutsche Forschungsgemeinschaft under grant cluster of excellence ct.qmat (EXC 2147, project-id 390858490).

\appendix

\section{\label{appsec:piecewiselin}Other forms of the gauge potential}

\begin{figure}[h]
    \includegraphics[width=0.9\columnwidth]{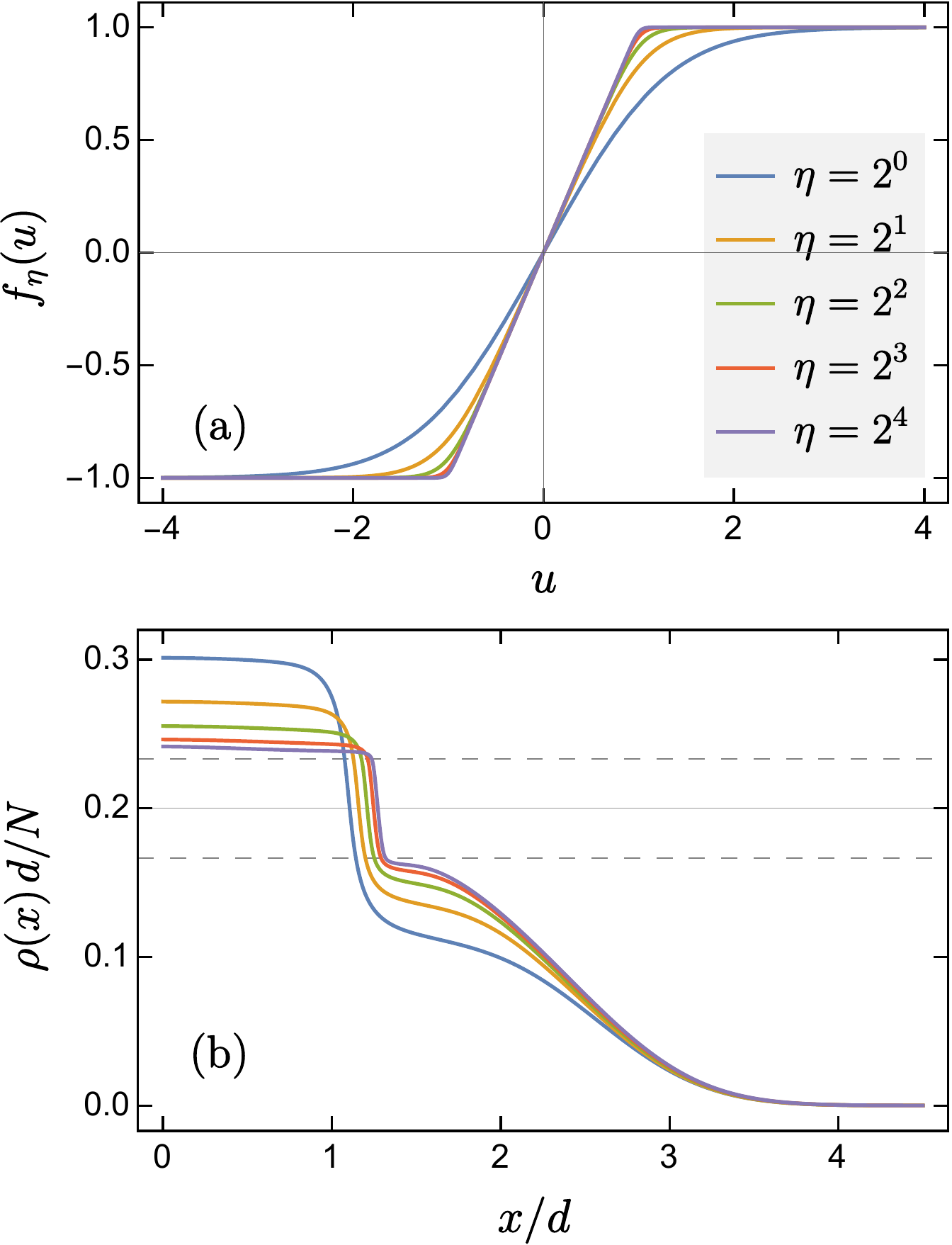}
    \centering
    \caption{(a) One-parameter family of curves that yield a piecewise linear variation of the gauge potential for $\eta \to \infty$. (b) Resulting ground-state density profiles for $\tilde{g} = 40$, $N l / d = 30$, $k_0 d = 3$, and $\rho_c d / N = 0.2$ (solid horizontal line), corresponding to the red curve in Fig.~\ref{fig:profiles}(b), with the same color convention as in (a). As $\mathbfcal{A}(\rho)$ becomes sharper the domain wall gets more pronounced between $\rho_c \pm 1/l$ (dashed horizontal lines).
    }
    \label{appfig:piecewise}
\end{figure}

Our predictions for the domain wall do not rely on the $\tanh$ variation of the gauge potential [Eq.~\eqref{eq:tanh}]. To illustrate this point we consider a different set of vector potentials $\mathbfcal{A} = \hbar k_0 f_{\eta} [(\rho - \rho_c) l] \push \hat{\mathbf{x}}$, where
\begin{equation}
    f_{\eta} (u) \coloneqq 
    \frac{1}{2\eta} \ln \pull \left[ 
    \frac{\cosh(\eta(1+u))}{\cosh(\eta(1-u))} 
    \right].
    \label{appeq:picewise}
\end{equation}
These functions are motivated by requiring their slope to reproduce the unit box function for $\eta \to \infty$, $f_{\eta}^{\prime}(u) = (1/2) \smash{\big[ \pull \tanh(\eta(u+1)) - \tanh(\eta(u-1)) \big]}$. Thus, for $\eta \lesssim 1$, $f_{\eta}(u)$ is a smooth ramp, whereas for $\eta \gg 1$ it assumes a piecewise linear form, as shown in Fig.~\ref{appfig:piecewise}(a). Figure~\ref{appfig:piecewise}(b) shows that as $\eta$ is increased the domain wall becomes more clearly confined between $\rho = \rho_c \pm 1/l$ while its slope $\rho^{\prime}(x)$ is unaltered. From Eq.~\eqref{eq:stationarydensity} the curvature $\rho^{\prime\prime}(x)$ at an edge of the domain wall is limited by $k_0^2 l \rho_c^2$.

\section{\label{appsec:slope}Density gradient at the domain wall}

\begin{figure}[h]
\includegraphics[width=0.9\columnwidth]{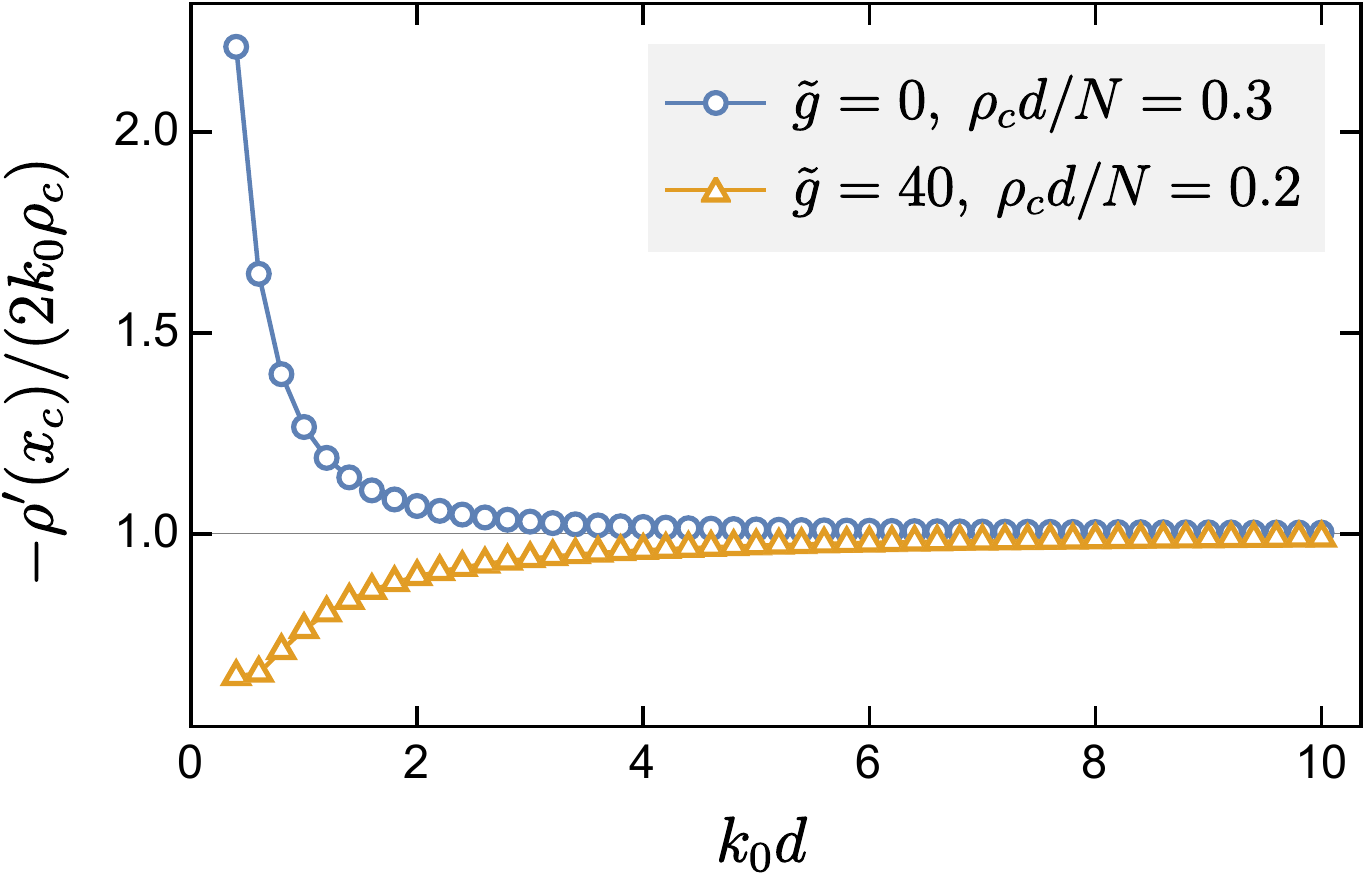}
    \centering
    \caption{
    Slope at the domain wall location $x_c$, where $\rho(x_c) \coloneqq \rho_c$, for $Nl/d=30$, corresponding to Figs.~\ref{fig:profiles}(a) (blue) and ~\ref{fig:profiles}(b) (orange). In both cases $\rho^{\prime}(x_c)$ approaches $-2k_0 \rho_c$ at large $k_0$, in accordance with our estimate from Eq.~\eqref{eq:domainwallenergy}.  
    }
    \label{appfig:slope}
\end{figure}

In Fig.~\ref{appfig:slope} we plot the numerically obtained density gradient at the domain wall for the profiles in Figs.~\ref{fig:profiles}(a,b). As the gauge potential increases, the slope converges to our estimate from the domain wall energy.

\section{\label{appsec:criticalpoint}Variation across the phase transition}

\begin{figure}
    \includegraphics[width=0.9\columnwidth]{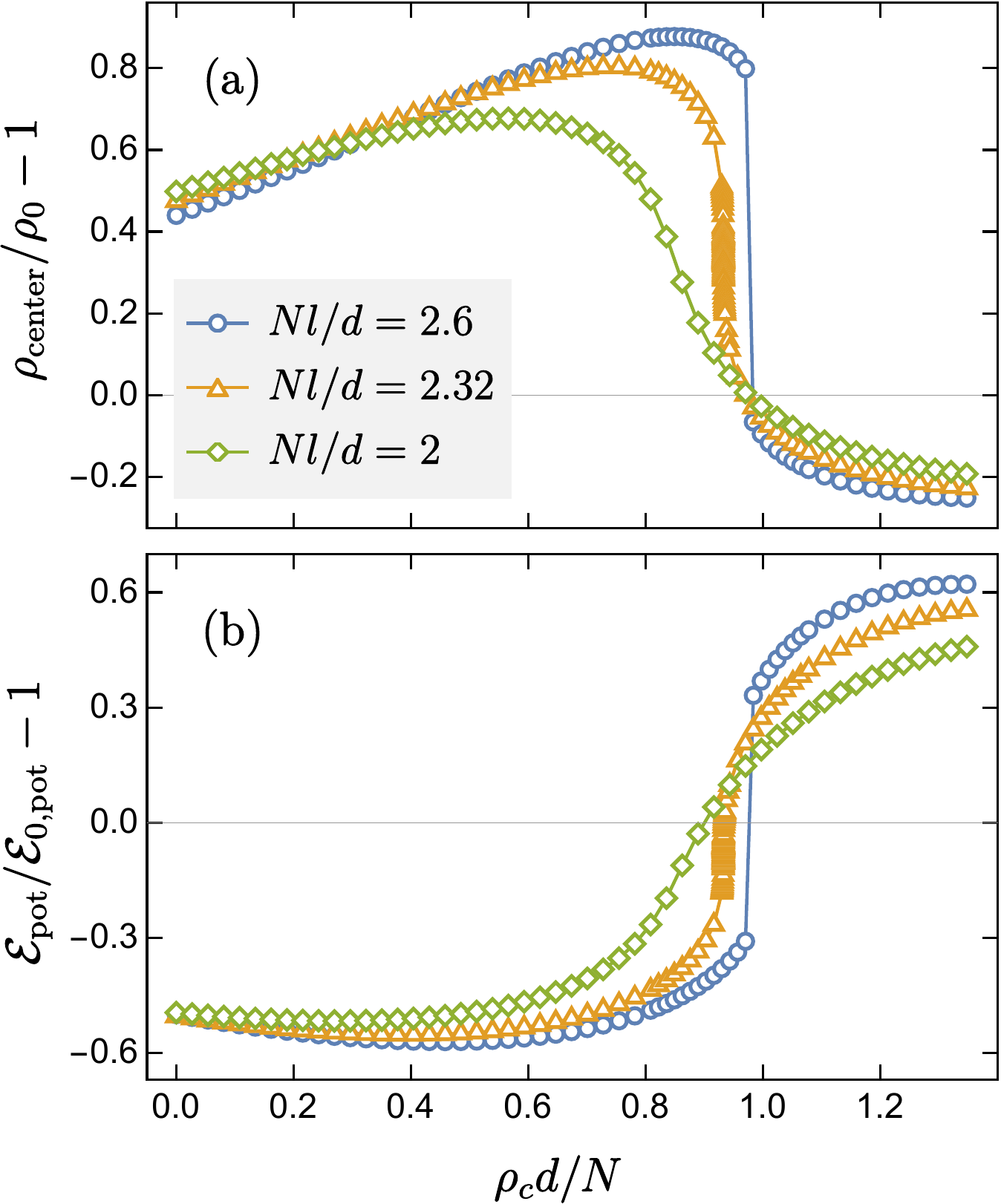}
    \centering
    \caption{
    Variation of the (a) peak density and (b) potential energy, measured relative to the unperturbed ground state, across the phase diagram for $\tilde{g}=10$ and $k_0 d = 2$ shown in Fig.~\ref{fig:phasetran}(b). The observables exhibit an infinite slope at the critical point $l_c$ (orange), a jump across a first-order transition for $l > l_c$ (blue), and a smooth crossover for $l < l_c$ (green).
    }
    \label{appfig:critical}
\end{figure}

Figure~\ref{appfig:critical} shows how ground-state observables change across the phase transition in Fig.~\ref{fig:phasetran}(b). As the smoothness parameter $l$ is decreased, a discontinuous jump turns into an infinite slope at the critical point and subsequently becomes a crossover.

\pagebreak

\bibliography{refs}

\end{document}